\documentclass[prl,twocolumn,twoside,preprintnumbers,superscriptaddress,showpacs,showkeys]{revtex4}
\usepackage{amsmath,slashed}
\usepackage{graphicx,graphics}
\usepackage{dcolumn}
\usepackage[hyperfootnotes=false]{hyperref}
\usepackage{xspace}

\newcommand{\mpi}{M_{\pi}}
\newcommand{\Order}{\mathcal{O}}

\newcommand{\GeV}{\,\text{GeV}}
\newcommand{\BR}{\text{BR}}
\newcommand{\beq}{\begin{equation}}
\newcommand{\eeq}{\end{equation}}

\renewcommand{\Im}{\text{Im}\,}

\begin{document}

\preprint{INT-PUB-17-005, CERN-TH-2017-014, NSF-KITP-17-012}
\title{Rescattering effects in the hadronic-light-by-light contribution \\[1mm] to the anomalous magnetic moment of the muon}

\author{Gilberto Colangelo}
\affiliation{Albert Einstein Center for Fundamental Physics, Institute for Theoretical Physics,
University of Bern, Sidlerstrasse 5, 3012 Bern, Switzerland }
\author{Martin Hoferichter}
\affiliation{Institute for Nuclear Theory, University of Washington, Seattle, WA 98195-1550, USA}
\affiliation{Kavli Institute for Theoretical Physics, University of California, Santa Barbara, CA 93106, USA}
\author{Massimiliano Procura\footnote[2]{On leave from the University of Vienna.}}
\affiliation{Theoretical Physics Department, CERN, Geneva, Switzerland}
\author{Peter Stoffer}
\affiliation{Helmholtz-Institut f\"ur Strahlen- und Kernphysik (Theory) and Bethe Center for Theoretical Physics, University of Bonn, 53115 Bonn, Germany}
\affiliation{Department of Physics, University of California at San Diego, La Jolla, CA 92093, USA}

\begin{abstract}
We present a first model-independent calculation of $\pi\pi$ intermediate
states in the hadronic-light-by-light (HLbL) contribution to the anomalous
magnetic moment of the muon  $(g-2)_\mu$ that goes beyond the scalar QED
pion loop. To this end we combine a recently developed dispersive
description of the HLbL tensor with a partial-wave expansion and
demonstrate that the known scalar-QED result is recovered after
partial-wave resummation. Using dispersive fits to high-statistics data for
the pion vector form factor, we provide an evaluation of the full pion
box, $a_\mu^{\pi\text{-box}}=-15.9(2)\times 10^{-11}$.
We then construct suitable input for the $\gamma^*\gamma^*\to\pi\pi$
helicity partial waves based on a pion-pole left-hand cut and show that for
the dominant charged-pion contribution this representation is consistent
with the two-loop chiral prediction and the COMPASS measurement for the pion
polarizability. This allows us to reliably estimate $S$-wave rescattering
effects to the full pion box 
and leads to our final estimate for the sum of these two contributions:
$a_\mu^{\pi\text{-box}} + a_{\mu,J=0}^{\pi\pi,\pi\text{-pole 
    LHC}}=-24(1)\times 10^{-11}$.
\end{abstract}


\maketitle

\section{Introduction}

The final report of the BNL E821 experiment~\cite{Bennett:2006fi} dominates
the world average for the experimental value of the anomalous magnetic
moment of the muon $(g-2)_\mu$, establishing a departure from its
Standard-Model (SM) expectation by about $3\sigma$ and thus providing an
intriguing hint for New Physics that makes the improved measurement at FNAL
E989~\cite{Grange:2015fou} as well as a potential independent determination
at J-PARC E34~\cite{Saito:2012zz} highly anticipated
(see~\cite{Gorringe:2015cma} for a detailed comparison of the two methods).
However, the significance of the deviation crucially depends on the details
of the SM evaluation. Even more so, a sound interpretation of the future
experiments demands that also the theory uncertainties be carefully
reassessed and ideally reduced in parallel with the experimental
improvement.

The by far dominant uncertainties in the SM prediction arise from hadronic
contributions: hadronic vacuum polarization (HVP) at second order in the
fine structure constant $\alpha$ and hadronic light-by-light scattering
(HLbL) at $\Order(\alpha^3)$~\cite{Jegerlehner:2009ry}. With higher-order
iterations of the same topologies already under good
control~\cite{Calmet:1976kd,Hagiwara:2011af,Kurz:2014wya,Colangelo:2014qya}
most theoretical efforts are concentrated on reducing the uncertainties in
the calculation of the HVP and HLbL contributions.
But while analyticity and unitarity allow one to express the former
in terms of
$\sigma(e^+e^-\to\text{hadrons})$~\cite{Bouchiat:1961,Blum:2013xva},
which is well measured, an expression of the HLbL contribution in terms of
measurable quantities was not known until recently. So, traditionally, HLbL
scattering has been estimated using hadronic models relying on different
limits of QCD---large-$N_c$, chiral symmetry, perturbative expansion---as
guiding
principles~\cite{deRafael:1993za,Bijnens:1995cc,Bijnens:1995xf,Bijnens:2001cq,Hayakawa:1995ps,Hayakawa:1996ki,Hayakawa:1997rq,Pivovarov:2001mw,Knecht:2001qg,Knecht:2001qf,RamseyMusolf:2002cy,Melnikov:2003xd,Erler:2006vu,Goecke:2010if,Greynat:2012ww},
which, however, complicates the assessment of the theoretical uncertainty
as well as the identification of strategies for systematic improvements,
making it emerge as a potential
roadblock~\cite{Prades:2009tw,Benayoun:2014tra}. 

In a series of recent
papers~\cite{Hoferichter:2013ama,Colangelo:2014dfa,Colangelo:2014pva,Stoffer:2014rka,Colangelo:2015ama}
we have shown that also the HLbL contribution can be expressed in terms of
measurable quantities, albeit not in a form as compact as for HVP. In our
model-independent approach based on dispersion relations, we have organized
the calculation of the HLbL tensor in terms of its singularities, i.e.\
single-particle poles and unitarity cuts, by expanding in the mass of
intermediate states~\cite{footnote}.
Individual terms in this expansion can be uniquely
defined in terms of form factors and scattering amplitudes, which, at least
in principle, are accessible to experiment. In this way, the notion of
pion-pole and pion-box contributions becomes unambiguous, and the first
terms in the expansion---pseudoscalar poles from $\pi^0$, $\eta$, $\eta'$
intermediate states---are fully determined by the corresponding
doubly-virtual transition form factors. Progress on the pseudoscalar-pole
contributions hinges on improved input for these form factors, in
combination with constraints on the asymptotic
behavior~\cite{Melnikov:2003xd}, and only concerns a few of
the scalar functions that are necessary for a full description of the HLbL
tensor. A program to reconstruct the transition form factors based on a
combination of unitarity, analyticity, and perturbative QCD with
experimental data is currently under
way~\cite{Stollenwerk:2011zz,Niecknig:2012sj,Schneider:2012ez,Hoferichter:2012pm,Hanhart:2013vba,Hoferichter:2014vra,Kubis:2015sga,Xiao:2015uva,Nyffeler:2016gnb}.

Next in the expansion are two-pion intermediate states. As demonstrated
in~\cite{Colangelo:2015ama}, the one-loop diagrams evaluated in scalar QED
(sQED), including pion vector form factors at each vertex to account for
the photon virtuality, provide an exact representation of the contribution
of two-pion intermediate states where only the pion-pole contribution to
the left-hand cut (LHC) of the $\gamma^*\gamma^*\to\pi\pi$ amplitudes is
retained. Thus the dispersive approach unambiguously defines the
gauge-invariant pion-box topology in terms of the pion vector form factor,
a very well measured quantity. Here, we present a numerical evaluation
of the pion box using a form factor fit to high-statistics data, in turn
using a dispersive representation to analytically continue the time-like
data into the space-like region required for the $(g-2)_\mu$ integral
and show that this contribution can be calculated with
negligible uncertainties. 

Extending our formalism beyond the pion box to account for two-pion
rescattering effects is not easy. Here we briefly review the technical
challenges, along with their solutions, to be faced when doing this extension,
and present a first numerical evaluation of $S$-wave 
$\pi\pi$-rescattering effects, which unitarize the pion-pole
contribution to $\gamma^*\gamma^*\to\pi\pi$. 
This constitutes the first step towards a full treatment of the
$\gamma^*\gamma^*\to\pi\pi$ partial
waves~\cite{GarciaMartin:2010cw,Hoferichter:2011wk,Moussallam:2013una}.
Our calculation settles the role of the pion polarizability, 
which enters at next-to-leading order in the chiral expansion of the HLbL
amplitude~\cite{Engel:2012xb,Engel:2013kda,Bijnens:2016hgx} and has been
suspected to produce sizable corrections in~\cite{Engel:2013kda}.  
In this paper, we illustrate the general strategy and present first
numerical results. Details of the formalism are relegated to~\cite{Colangelo:2017fiz}. 

\section{Dispersion relation for HLbL}

The central object in the calculation of the HLbL contribution to
$(g-2)_\mu$ is the hadronic four-point function
\begin{align}
	\Pi^{\mu\nu\lambda\sigma}(q_1,q_2,q_3)&= -i \int d^4x \, d^4y \, d^4z \, e^{-i(q_1 \cdot x + q_2 \cdot y + q_3 \cdot z)} \notag\\
	&\hspace{-20pt}\times\langle 0 | T \{ j_\text{em}^\mu(x) j_\text{em}^\nu(y) j_\text{em}^\lambda(z) j_\text{em}^\sigma(0) \} | 0 \rangle
\end{align}
of four electromagnetic currents
\begin{align}
	j_\text{em}^\mu = \bar q Q \gamma^\mu q, \quad Q = \text{diag}\left(\frac{2}{3}, -\frac{1}{3}, -\frac{1}{3}\right),
\end{align}
with momenta $q_i$ as indicated, $q_4=q_1+q_2+q_3$, and quark fields $q = (u , d, s)^T$. 

To be able to reconstruct the HLbL tensor $\Pi^{\mu\nu\lambda\sigma}$ with
dispersion relations, it is imperative to use a decomposition into scalar
functions that are free of kinematic singularities and zeros. Such a
representation can be obtained following the general recipe put forward by
Bardeen, Tung~\cite{Bardeen:1969aw}, and Tarrach~\cite{Tarrach:1975tu}
(BTT), resulting in 
\beq
\Pi^{\mu\nu\lambda\sigma} = \sum_{i=1}^{54} T_i^{\mu\nu\lambda\sigma} \Pi_i,
\eeq
with scalar functions $\Pi_i$ depending on the Mandelstam variables
$s=(q_1+q_2)^2$, $t=(q_1+q_3)^2$, $u=(q_2+q_3)^2$ as well as the
virtualities $q_i^2$, and Lorentz structures
$T_i^{\mu\nu\lambda\sigma}$~\cite{Stoffer:2014rka,Colangelo:2015ama}. This
decomposition fulfills gauge invariance manifestly 
\beq
\{q_1^\mu, q_2^\nu, q_3^\lambda, q_4^\sigma\} T^i_{\mu\nu\lambda\sigma}=0,
\eeq
is highly crossing symmetric (with only $7$ distinct structures, all
remaining 47 being related to these by crossing transformations), and
ensures that the coefficient functions $\Pi_i$ do not contain kinematic
singularities and zeros. 
In addition, the BTT decomposition typically allows for a very economical
representation of HLbL amplitudes, e.g.\ one of the structures coincides
with the amplitude for a pseudoscalar pole, while even the sQED amplitude
becomes very compact once expressed in terms of BTT
functions~\cite{Colangelo:2017fiz}. For the contribution to $(g-2)_\mu$ a
three-dimensional integral representation is
available~\cite{Colangelo:2017fiz} 
\begin{align}
\label{master}
a_\mu^\text{HLbL} &= \frac{\alpha^3}{432\pi^2} \int_0^\infty d\Sigma\,\Sigma^3 \int_0^1 dr\, r\sqrt{1-r^2} \int_0^{2\pi} d\phi \notag\\
&\times\sum_{i=1}^{12} T_i(Q_1,Q_2,Q_3) \bar\Pi_i(Q_1,Q_2,Q_3),
\end{align}
where the $T_i$ are known kernel functions, the $\bar\Pi_i$ suitable linear
combinations of the BTT $\Pi_i$, and the Euclidean momenta squared are given
by~\cite{Eichmann:2015nra} 
\begin{align}
Q_{1,2}^2 &= \frac{\Sigma}{3} \left( 1 - \frac{r}{2} \cos\phi \mp \frac{r}{2}\sqrt{3} \sin\phi \right), \notag\\
Q_3^2 &= \frac{\Sigma}{3} \left( 1 + r \cos\phi \right). 
\end{align}
There are only $6$ distinct functions $\bar\Pi_i$, the remaining ones are again related to these by crossing symmetry.
It suffices to calculate the $\bar\Pi_i$ in the kinematic limit where
$q_4\to 0$, the transition to $(g-2)_\mu$ then proceeds by means
of~\eqref{master}.

\subsection{Two-pion intermediate states}

\begin{figure}[t]
 \centering
 \includegraphics[width=\linewidth]{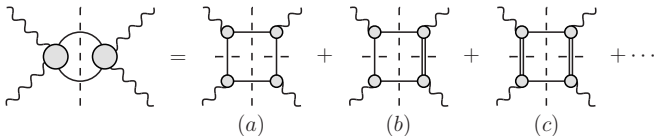}
 \caption{Two-pion-cut contributions to HLbL scattering. Solid/wiggly lines refer to pions/photons, respectively, while dashed lines indicate the cutting of propagators. Double lines generically denote heavier intermediate states, the gray blobs hadronic amplitudes. Crossed diagrams are omitted.}
 \label{fig:pipi}
\end{figure}

In a dispersive approach two-pion intermediate states comprise all
contributions that involve a two-pion cut, generically represented by the
left (unitarity) diagram in Fig.~\ref{fig:pipi}.  The dominant term is
obtained if in the $\gamma^*\gamma^*\to\pi\pi$ sub-amplitudes, in turn, the
pion is put on-shell, i.e.\ if the pion-pole contribution to the LHC is
isolated. In this case, the remaining hadronic amplitudes are given by pion
vector form factors, and as demonstrated in~\cite{Colangelo:2015ama}, this
class of two-pion intermediate states, the pure pion box in diagram $(a)$
in Fig.~\ref{fig:pipi}, reproduces the sQED pion loop with vertices
augmented by the appropriate pion form factors. The reason for this
behavior can be traced back to the fact that only the singularities of the
box diagrams in sQED matter, while the triangle and bulb diagrams are
simply required to restore gauge invariance.  Due to the high degree of
crossing symmetry, this pion-box contribution can be expressed in terms of
either fixed-$s$, -$t$, or -$u$ dispersion relations, or in a symmetrized
form
\begin{align}
\label{disp_box}
 \Pi_i^{\pi\text{-box}}(s,t,u) &= \frac{1}{3}
		\Bigg[ \frac{1}{\pi} \int_{4M_\pi^2}^\infty dt' \frac{\Im \Pi_i^{\pi\text{-box}}(s,t',u')}{t'-t}\\
		&\hspace{-50pt}+ \frac{1}{\pi} \int_{4M_\pi^2}^\infty du' \frac{\Im \Pi_i^{\pi\text{-box}}(s,t',u')}{u'-u}
		+\text{fixed-}t + \text{fixed-}u\Bigg]\notag.
\end{align}
In this case the representation is exact.

Once heavier intermediate states are considered, generically denoted by the
double lines in diagrams $(b)$ and $(c)$ in Fig.~\ref{fig:pipi}, a more
detailed investigation of the double spectral functions is required. In
practice, such contributions can be included using a partial-wave
expansion, in which case the sub-process becomes a polynomial in the
crossed variable and the crossed-channel cuts are neglected.    
Writing down all crossed
versions of the unitarity diagrams shown in Fig.~\ref{fig:pipi}, one sees
that each double spectral region appears exactly twice in a symmetrized
form as in~\eqref{disp_box}, so that the prefactor has to be changed from
$1/3\to 1/2$~\cite{Colangelo:2017fiz}, with corrections suppressed by the
mass scale of the neglected LHC.  In particular, this representation
becomes exact for $\pi\pi$-rescattering effects, which, by definition, are
polynomial in the crossed Mandelstam variable.

\subsection{Partial-wave expansion}
\label{sec:partial_waves}

Constraints from unitarity are most conveniently formulated in a
partial-wave expansion for HLbL helicity amplitudes
$h_{\lambda_1\lambda_2,\lambda_3\lambda_4}^J$ with angular momentum $J$ and
helicity labels $\lambda_i$. In this case the unitarity relation becomes
diagonal 
\beq
\label{unitarity}
\Im h_{\lambda_1\lambda_2,\lambda_3\lambda_4}^J(s)=\frac{\sigma_\pi(s)}{16\pi S}h_{J,\lambda_1\lambda_2}(s)h^*_{J,\lambda_3\lambda_4}(s),
\eeq
where $\sigma_\pi(s)=\sqrt{1-4\mpi^2/s}$ gives the phase space, $S=2$ a
symmetry factor in case of indistinguishable particles, and
$h_{J,\lambda_1\lambda_2}$ the helicity partial waves for
$\gamma^*\gamma^*\to\pi\pi$. Once formulated in isospin basis, Watson's
theorem~\cite{Watson:1954uc}, see also~\eqref{Watson} below, guarantees
that the phases on the right-hand side cancel to produce
a real imaginary part. 
The partial-wave expansion of the pion box is obtained if both $h_{J,\lambda_1\lambda_2}(s)$ and 
$h_{J,\lambda_3\lambda_4}(s)$ are identified with the partial-wave-projected Born terms, while the rescattering
effects correspond to the unitarity corrections to either subamplitude derived from~\eqref{Watson}. 

There are $41$ independent helicity amplitudes for the full HLbL
tensor, which reduce to $27$ if one photon is taken on-shell. Rewriting
the representation of the contribution to $(g-2)_\mu$,
Eq.~\eqref{master}, in such a way that only dispersive integrals over
imaginary parts of these $27$ helicity amplitudes appear is highly
nontrivial. By explicitly requiring that unphysical amplitudes drop out
in the final result, and that the two redundancies which appear in four space-time dimensions
$d=4$~\cite{Eichmann:2015nra} do not affect the result, one can derive a set 
of sum rules for the scalar functions. (In~\cite{Pascalutsa:2012pr}, sum rules for the 
special case of forward HLbL scattering have been derived.) 
These sum rules apply to the full amplitudes, but not necessarily at the
level of the partial-wave-expanded ones, producing an apparent dependence
on unphysical amplitudes that would only disappear after a resummation of
all partial waves. 

To avoid such pathologies, we were able to construct a set of $27$
amplitudes $\check\Pi_i$ related to the $27$ singly-on-shell
helicity amplitudes by a basis change that we have derived in explicit
analytic form. In the limit $q_4\to 0$ a subset of the $\check\Pi_i$
includes all the scalar functions needed as input
in~\eqref{master}~\cite{Colangelo:2017fiz}. Moreover, this set of $27$
amplitudes is manifestly free of Tarrach~\cite{Tarrach:1975tu} or $d=4$
ambiguities~\cite{Eichmann:2015nra}. For singly-on-shell kinematics,
there still exist $15$ sum rules among the $27$ helicity amplitudes, which
we have exploited to optimize to a certain degree the representation
with respect to the convergence of the partial-wave expansion. 
This formalism is now ready to be applied to the evaluation of rescattering
effects, but before doing that we test it with the help of the pion box and
study how well we are able to reproduce its numerical value by resumming
the partial-wave expansion.  

\section{Pion box}

The formalism for dealing with the pion box has been developed in~\cite{Colangelo:2015ama}. 
Here we provide a first numerical evaluation thereof
with a realistic pion form factor. The latter has been
obtained by fitting a dispersive representation as suggested
in~\cite{Leutwyler:2002hm,Colangelo:2003yw} to both
space-like~\cite{Amendolia:1986wj} and
time-like~\cite{Achasov:2006vp,Akhmetshin:2006bx,Aubert:2009ad,Ambrosino:2010bv,Babusci:2012rp,Ablikim:2015orh}
form factor data (similar representations have been used before
in~\cite{DeTroconiz:2001rip,deTroconiz:2004yzs,Ananthanarayan:2013zua,Ananthanarayan:2016mns,Hoferichter:2016duk,Hanhart:2016pcd}),
with the result 
\beq
\label{eq:pibox}
a_\mu^{\pi\text{-box}}=-15.9(2)\times 10^{-11},
\eeq
and an uncertainty determined from the differences between the time-like
data sets as well as the details of the fit representation. 
The main reduction in uncertainty compared to earlier evaluations of a ``pion loop''~\cite{Bijnens:1995cc,Hayakawa:1995ps} 
is due to the insight that the pion box, defined as two-pion intermediate states with a pion-pole left-hand cut, is the 
unambiguous first term in the expansion and can be expressed in terms of a hadronic observable,
the pion vector form factor, which is very well known phenomenologically.

The pion box also provides an ideal test case for the framework presented
in the previous section since the full result is known and explicit
expressions for all BTT scalar functions are available. As a first step, we
verified that the sum rules encountered in the context of the partial-wave
expansion are fulfilled. Second, in the special case of the pion box a
fixed-$s$, -$t$, -$u$ representation should each hold, combining to the
symmetrized version in~\eqref{disp_box}, so that the convergence can be
studied in each channel separately. The results, for simplicity obtained by
using a vector-meson-dominance pion form factor
$F_\pi^V(q^2)=M_\rho^2/(M_\rho^2-q^2)$, with $a_\mu^{\pi\text{-box,
    VMD}}=-16.4\times 10^{-11}$, are shown in Table~\ref{tab:sQED},
demonstrating that each representation approaches the full result (going up
to $J_\text{max}=20$, we checked that also the remaining differences
disappear after partial-wave resummation). 
The vanishing $S$-wave contribution for fixed-$s$ is well
understood and partly a matter of convention in the choice of the $6$
functions $\bar\Pi_i$, see~\cite{Colangelo:2017fiz}.
In concrete applications, the prescription of changing the prefactor
in~\eqref{disp_box} as explained above combines the three representations
in a way that best captures the physics (such as a resonance) in all
channels at once, which means that the convergence patterns for fixed-$t$
or -$u$  are more representative of realistic cases and the average of the
three should be viewed as a worst-case scenario. But even that displays a
very reasonable convergence behavior.  

\begin{table}[t]
\renewcommand{\arraystretch}{1.3}
\centering
\begin{tabular}{crrrr}\toprule
$J_\text{max}$ & fixed-$s$ & fixed-$t$ & fixed-$u$ & average\\\colrule
$0$ & $0.0\%$ & $106.2\%$ & $106.2\%$ & $70.8\%$\\
$2$ & $73.9\%$ & $102.3\%$ & $92.7\%$ & $89.6\%$\\
$4$ & $89.2\%$ & $101.5\%$ & $96.4\%$ & $95.7\%$\\
$6$ & $94.3\%$ & $100.7\%$ & $97.9\%$ & $97.6\%$\\
$8$ & $96.5\%$ & $100.4\%$ & $98.7\%$ & $98.5\%$\\
\botrule
\end{tabular}
\caption{Saturation of $a_\mu^{\pi\text{-box}}$ for maximal angular momentum $J_\text{max}$.}
\label{tab:sQED}
\end{table}

\section{$\boldsymbol{\pi\pi}$ rescattering effects}

We now turn to the evaluation of rescattering effects, as a first
important step to go beyond the pion-box contribution. The helicity
amplitudes $h_{J,\lambda_1\lambda_2}(s)$ entering~\eqref{unitarity}, satisfy themselves a unitarity relation
\beq
\label{Watson}
\Im h^I_{J,\lambda_1\lambda_2}(s) = \sin \delta_J^{I}(s) e^{-i\delta_J^I(s)} h^I_{J,\lambda_1\lambda_2}(s),
\eeq
with isospin labels $I$ and $\pi\pi$ phase shifts $\delta_J^{I}$. This
relation is clearly violated for the (real) Born terms alone, but this
deficiency can be easily repaired by solving the dispersion relation for
the subprocess  $\gamma^*\gamma^*\to\pi\pi$. 

In contrast to the on-shell and singly-virtual
case~\cite{GarciaMartin:2010cw,Hoferichter:2011wk,Moussallam:2013una}, the
calculation of the $\gamma^*\gamma^*\to\pi\pi$ partial waves for two
off-shell photons is complicated by the fact that even for $S$-waves two
different helicity partial waves, $h_{0,++}$ and $h_{0,00}$, become
coupled, including off-diagonal kernel functions required to eliminate
kinematic singularities~\cite{Colangelo:2014dfa,Colangelo:2015ama}. Here,
we apply this framework to construct the $\gamma^*\gamma^*\to\pi\pi$
amplitudes that correspond to the rescattering corrections to the Born
terms, whose solution can still be derived based on Muskhelishvili--Omn\`es
methods~\cite{Muskhelishvili:1953,Omnes:1958hv}. We use $\pi\pi$ phase
shifts based on the modified inverse-amplitude
method~\cite{GomezNicola:2007qj}, for the main reason that it has a
  simple analytic expression which is convenient to use in combination with
  Muskhelishvili--Omn\`es methods, while at the same time it reproduces
accurately the low-energy properties of the phase shifts as well as pole 
position and couplings of the $f_0(500)$ resonance. This phase shift
departs from the correct one just below the $K\bar K$ threshold because it
does not feature the sharp rise due to the $f_0(980)$ resonance but
continues flat with a smooth high-energy behavior. A full-fledged
evaluation of the $f_0(980)$ resonance would require a proper treatment of
the $K\bar K$ channel, which is beyond the scope of this first estimate. We
can, on the other hand, test the sensitivity to the asymptotic part of the
dispersive integrals by studying solutions with different cutoff values
$\Lambda=[1\GeV,\infty)$, constructed with finite-matching-point
techniques~\cite{Buettiker:2003pp,Hoferichter:2011wk,Ditsche:2012fv,Hoferichter:2012wf,Hoferichter:2015hva}.
Moreover, we checked that for low values of $\Lambda$ phase shifts obtained
by solving Roy
equations~\cite{Colangelo:2001df,Caprini:2011ky,GarciaMartin:2011cn} lead
to equivalent results. 

\begin{table}[t]
\renewcommand{\arraystretch}{1.3}
\centering
\begin{tabular}{crrrr}\toprule
cutoff & $1\GeV$ & $1.5\GeV$ & $2\GeV$ & $\infty$\\\colrule
$I=0$ & $-9.2$ & $-9.5$ & $-9.3$ & $-8.8$\\
$I=2$ & $2.0$ & $1.3$ & $1.1$ & $0.9$\\
sum & $-7.3$ & $-8.3$ & $-8.3$ & $-7.9$\\
\botrule
\end{tabular}
\caption{$S$-wave rescattering corrections to $a_\mu^{\pi\text{-box}}$, in units of $10^{-11}$, for both isospin components and in total.}
\label{tab:rescatt}
\end{table}

The results for the rescattering contribution, summarized in
Table~\ref{tab:rescatt}, are indeed stable over a wide range of cutoffs,
indicating that our input for the $\gamma^*\gamma^*\to\pi\pi$ partial waves
reliably unitarizes the Born-term LHC, which should indeed dominate at low
energies. In addition, we checked that the only sum rule that receives
$S$-wave contributions is already saturated at better than $90\%$,
completely in line with the expectation that the sum rules will be
fulfilled only after partial-wave resummation. The isospin-$0$ part of the
result can be interpreted as a model-independent implementation of the
contribution from  
the $f_0(500)$ of about
$-9\times 10^{-11}$ to HLbL scattering in $(g-2)_\mu$. In total, we obtain
for the $\pi\pi$-rescattering effects related to the pion-pole LHC 
\beq
\label{amupipi}
a_{\mu,J=0}^{\pi\pi,\pi\text{-pole LHC}}=-8(1)\times 10^{-11},
\eeq
where the error is dominated by the uncertainties related to the asymptotic
parts of the integral, see Table~\ref{tab:rescatt}.
Improving the energy region $\gtrsim 1\GeV$ requires the inclusion of 
the $K \bar K$ channel as well as higher contributions to the LHC, 
neither of which can be expressed in terms of the pion vector form
factor. Very likely, such effects beyond pion states will be less precisely estimated.

Finally, it is instructive to consider the separate contributions not in
the isospin, but in the charge basis. In this case, the unitarity
relation~\eqref{unitarity} is no longer diagonal and it is not possible
to define unambiguously the contribution of each of the charge states.
Irrespective of the detailed convention for the separation, charged-pion
states are expected to strongly dominate, e.g.\ in the chiral expansion
neutral-pion intermediate states first appear at three-loop order.  
The derivative of the Born-term-subtracted amplitude $h_{0,++}(s)$ is
related to the pion dipole polarizability $\alpha_1-\beta_1$, to which the
unitarized pion-pole LHC contributes  
\begin{align}
 (\alpha_1-\beta_1)^{\pi^\pm,\pi\text{-pole LHC}}&=(5.4\ldots 5.8)\times
 10^{-4}\,\text{fm}^3, \notag \\
 (\alpha_1-\beta_1)^{\pi^0,\pi\text{-pole LHC}}&=(11.2\ldots 8.9)\times
 10^{-4}\,\text{fm}^3 ,
\end{align}
for $\Lambda=1\GeV\ldots\infty$. For the charged pion this result is in
perfect agreement with the chiral $2$-loop prediction
$5.7(1.0)$~\cite{Gasser:2006qa} (in the same units) as well as the recent
COMPASS measurement
$4.0(1.2)_\text{stat}(1.4)_\text{syst}$~\cite{Adolph:2014kgj}. In contrast,
the $2$-loop prediction for the neutral pion,
$-1.9(0.2)$~\cite{Gasser:2005ud}, is substantially smaller in size and
  has the opposite sign of what we get from our representation. 
This failure, however, is not reason for much concern because we are
not yet including here the main contributions to the LHC of the amplitude
for neutral pions, i.e.\ vector-meson exchange involving
$V=\omega,\rho$. Due to the scaling with $\Gamma_{V\to\pi\gamma}$, the
relative impact on the neutral channel~\cite{Olive:2016xmw} 
\beq
\frac{\Gamma_\omega\times \BR[\omega\to\pi^0\gamma]+\Gamma_\rho \times
  \BR[\rho^0\to\pi^0\gamma]}{\Gamma_\rho \times \BR[\rho^\pm\to\pi^\pm\gamma]}\sim
12 
\eeq
is an order of magnitude larger, so that heavier intermediate states allow
one to repair $(\alpha_1-\beta_1)^{\pi^0}$ without spoiling agreement in
the charged channel. In summary, the rescattering effects
in~\eqref{amupipi} are dominated by the charged pion, with input for the
$\gamma^*\gamma^*\to\pi\pi$ partial waves fully consistent with its dipole
polarizability. For this reason~\eqref{amupipi} can be considered a
model-independent implementation of effects related to the low-energy
constants $L_9$ and $L_{10}$, which were suspected to produce large effects
in~\cite{Engel:2013kda}. Our calculation proves 
that this is not the case, and that the related rescattering corrections
are indeed of very reasonable size (a similar conclusion was reached
within a model approach in~\cite{Bijnens:2016hgx}). In this context it
should be stressed that our analysis does not rely on chiral operators,
thus avoiding the pathologies in their high-energy behavior and the need to
cure them. The polarizabilities enter here as the limit of our
$\gamma^*\gamma^*\to\pi\pi$ amplitudes at a particular kinematic point that
does not contribute to the dispersive integrals directly, providing an
important cross check of the low-energy limit.

In conclusion, we have shown that our framework allows us to
estimate very accurately the combined effect of two-pion intermediate
states generated by a pion-pole LHC and its $S$-wave unitarization 
\beq
a_\mu^{\pi\text{-box}} + a_{\mu,J=0}^{\pi\pi,\pi\text{-pole
    LHC}}=-24(1)\times 10^{-11}, 
\eeq 
which is considered to be among the most important contributions after the
dominant pseudoscalar poles, but was so far affected by significant
uncertainties. This first numerical result based on the dispersive approach
lays the foundation for extensions towards higher partial waves, an
improved LHC in the $\gamma^*\gamma^*\to\pi\pi$ subamplitudes as
well as higher-mass intermediate states, all important prerequisites for a
model-independent evaluation of the complete HLbL contribution to
$(g-2)_\mu$.  

\section*{Acknowledgments}
\begin{acknowledgments}
Financial support by
the DFG (SFB/TR 16, ``Subnuclear Structure of Matter,'' SFB/TR 110,
``Symmetries and the Emergence of Structure in QCD''),   
the DOE (Grant No.\ DE-FG02-00ER41132), 
the National Science Foundation (Grant No.\ NSF PHY-1125915),
and the Swiss National Science Foundation
is gratefully acknowledged.
M.P.\ is supported by a Marie Curie Intra-European Fellowship of the
European Community's 7th Framework Programme under contract number
PIEF-GA-2013-622527
and P.S.\ by a grant of the Swiss National Science Foundation (Project No.\ P300P2\_167751).
\end{acknowledgments}

\end{document}